\documentclass[a4paper,usenatbib]{mnras}

\usepackage{mathptmx}

\usepackage[T1]{fontenc}
\usepackage{ae,aecompl}

\usepackage{amsmath}	
\usepackage{amssymb}	
\usepackage{graphicx}	

\title[Physical connection]{Scale Invariance, Horizons, and Inflation}
\author[Maeder \& Gueorguiev]{
Andre Maeder $^{1}$\thanks{E-mail: Andre.Maeder at UniGe.ch}
and Vesselin G. Gueorguiev $^{2,3}$\thanks{E-mail: Vesselin at MailAPS.org}\\
$^{1}${Geneva Observatory - chemin des Maillettes 51, CH-1290 Sauverny, Switzerland}\\
$^{2}${Institute for Advanced Physical Studies, Sofia, Bulgaria}\\
$^{3}${Ronin Institute for Independent Scholarship, Montclair, NJ, USA}
}

\date{Accepted XXX. Received YYY; in original form ZZZ}

\pubyear{2019}

\begin{document}
\label{firstpage}
\pagerange{\pageref{firstpage}--\pageref{lastpage}}
\maketitle

\begin{abstract}
Maxwell equations and the  equations of General Relativity are scale invariant in  empty space.
The presence of charge or currents in electromagnetism 
or the presence of matter in cosmology are preventing scale invariance. 
The question arises on how much matter within the horizon is necessary to kill scale invariance. 
The scale invariant field equation, first written by Dirac in 1973 
and then revisited by Canuto et al. in 1977, 
provides the starting point to address this question.  
The resulting cosmological models show that,  as soon as matter  is present, 
the effects of scale invariance rapidly  decline from $\varrho=0$ to $\varrho_{\mathrm{c}}$,
and are  forbidden for  densities above $\varrho_{\mathrm{c}}$.
The absence of scale invariance  in this case  is  consistent with considerations about causal connection. 
Below  $\varrho_{\mathrm{c}}$, scale invariance appears as an open possibility,
which also depends on the occurrence of inflation  in the scale invariant context. 
In the  present approach, we identify the scalar field of the empty space in the Scale Invariant Vacuum (SIV) context
to the scalar field $\varphi$ in  the energy density 
$\varrho  =   \frac{1}{2} \dot{\varphi}^2 + V(\varphi)$ of the vacuum at inflation. 
This leads to some constraints on the potential. 
This identification also solves the so-called ``cosmological constant problem''. 
In the framework of scale invariance,
an inflation with a large number of e-foldings is also predicted. We conclude that 
scale invariance for models  with densities below $\varrho_{\mathrm{c}}$ is an open possibility; the final answer 
may come from high redshift observations, where differences from the $\Lambda$CDM models appear.

\end{abstract}


\section{Introduction} \label{intro}

The laws of physics are generally not unchanged under a
change of scale, a fact discovered by Galileo Galilei, as pointed out by \citet{Feynman63}.
Feynman also emphasized that  the scale references are closely related to the material
content of the medium. An empty Universe would be scale invariant, 
since there would be nothing to  define a  scale. 
Indeed,  the Maxwell equations in absence of charge and currents are scale invariant.
Similarly, the field equation of General Relativity (GR), without cosmological constant,
is scale invariant for empty space\citep{Bondi90}, a fact that is rarely mentioned.

Concerning Cosmology and the evolution of the Universe, 
it is however not so clear which amount of matter is necessary to kill scale invariance.
Would a single atom in the whole Universe be sufficient to define scales throughout    at any time?
The  problem of scales is related to the existence of physical connection and causality,
which relates  different regions of the Universe. In this context,
the horizons and the inflation play an important role.
Here, we examine  the domain of cosmological parameters,  
for which physical connection exists and likely enforces
well defined physical scales. For some cosmological conditions, there may be no causality 
connection and thus  the door could be open to scale invariance. 
The question is closely  related to the problems of  horizons and  inflation.

In Section \ref{mod}, we first follow our main objective, 
{\it{i.e.}}  to know what amount matter is killing scale invariance,
on the basis of the scale invariant field equations established by \citet{Dirac73} and  \citet{Canuto77}, 
as well from the cosmological  solutions of these equations.
In Section \ref{horiz}, we examine  the problem in terms of the particle- and event- horizons. We suggest that
another definition, the physical horizon, is more meaningful and places constraints on the mean density of the Universe.
In Section \ref{infl}, we study the scalar field associated to scale invariance in relation with 
the scalar field of the inflation.  We show that the scale invariant context and equations  
permits the occurrence of an  inflation with a large number of e-foldings.
The conclusions and perspectives are given in Section \ref{concl}.

\section{The answer for SIV cosmology models}  \label{mod}
We first explore to what extent the existing studies on scale invariance
provide some information on the above mentioned problem: what amount of matter is killing scale invariance?
\citet{Dirac73} and  \citet{Canuto77} have established the basis of scale invariant cosmology.
In addition to the general covariance
of General Relativity, the field equations are also invariant
upon the scale transformation of the form:
\begin{equation}
ds' \, = \, \lambda(x^{\mu}) \, ds  \,.
\label{lambda}
\end{equation}
\noindent
There, $ds'$ is the line element in GR and $ds$ in the scale invariant space, 
which is that of  Weyl's Integrable Geometry \citep{Weyl23,Dirac73}. 
We will see below (Sect. \ref{scal})  that the space-time is thus endowed 
with a scalar field  $\psi$ related to the above $\lambda(t)$.

\subsection{Brief recalls on the scale invariant cosmology} \label{recall}

We limit the recalls to  the necessary minimum to follow the developments below.
More details have been summarized by \citet{Canuto77} and \citet{MaedGueor20a}.
Scale invariant first and second derivatives, scale invariant Christoffel symbols,
Riemann-Christoffel tensor and total curvature have been obtained. 
By using the convention that primed quantities are expressions based on the 
line element in GR, while quantities without prime 
are the expressions related to the Weyl's Integrable Geometry and the 
corresponding relationship for the Ricci tensor \citep{Maeder17a}, we write:
\begin{eqnarray}
R_{\mu \nu}=R'_{\mu \nu}  - \kappa_{\mu ;\nu} - \kappa_{ \nu ;\mu} -2 \kappa_{\mu} \kappa_ {\nu} 
+ 2 g_{\mu \nu}\kappa^{ \alpha} \kappa_{ \alpha}- g_{\mu \nu} \kappa^{ \alpha}_{;\alpha}\\
R=R'+6\kappa^{ \alpha} \kappa_{ \alpha} - 6\kappa^{ \alpha}_{;\alpha}
\end{eqnarray}
The usual Einstein equations  can be generalized to the general scale invariant field equation 
\citep{Dirac73,Canuto77}:
\begin{equation} 
R_{\mu \nu} - \frac{1}{2} \ g_{\mu \nu} R = -8 \pi G T_{\mu \nu} -  \Lambda \, g_{\mu \nu} \, .
\end{equation}     
or explicitly
\begin{eqnarray}
R'_{\mu \nu} - \frac{1}{2} \ g_{\mu \nu} R' - \kappa_{\mu ;\nu} - \kappa_{ \nu ;\mu}
-2 \kappa_{\mu} \kappa_ {\nu} + 2 g_{\mu \nu} \kappa^{ \alpha}_{;\alpha}
- g_{\mu \nu}\kappa^{ \alpha} \kappa_{ \alpha} =  \nonumber \\
-8 \pi G T_{\mu \nu} - \lambda^2 \Lambda_{\mathrm{E}} \, g_{\mu \nu} \, .
\label{field}
\end{eqnarray}

\noindent
This field equation is the generalization of Einstein equation to also account for scale invariance in addition to general covariance.
We have in the scale invariant framework $\Lambda = \lambda^2 \Lambda_{\mathrm{E}}$.
$G$ is the gravitational constant, taken  as a true  constant. The above equation
contains additional terms depending on  the metrical connection $\kappa_{\nu}$, 
related to the scale factor $\lambda$, 
\begin{equation}
\kappa_{\nu} \,  = \,
- \frac{\partial \ln \lambda}{\partial x^{\nu}} \, .
\label{ka}
\end{equation}     
\noindent
It is immediately seen that two successive scale transformations  $\lambda=\lambda^{(1)}\lambda^{(2)}$ result in additive 
expression for the corresponding metrical connection $\kappa_{\nu}=\kappa^{(1)}_{\nu}+\kappa^{(2)}_{\nu}$.
For reasons of homogeneity and isotropy the scale factor  $\lambda$ in cosmology should depend on time only \citep{Maeder17a}, 
so that the only component of $\kappa_{\nu}$ is $\kappa_0$. We have in particular,
\[
\kappa_{\mu, \nu}= \kappa_{0, 0} = \frac{d \kappa_0}{dt} =\dot{\kappa}_0=-\frac{\dot{\lambda}}{\lambda} \,.
\]
In Weyl's Integrable Geometry, $\kappa_{\nu}$ is playing a fundamental role alike the $g_{\mu \nu}$. 
If $\lambda$ is a constant, one is brought back to the usual equations of GR.

With the FLWR parametrized metric, one is  lead to the following  
differential equations  for cosmological models \citep{Canuto77},
\begin{eqnarray}
\frac{8 \, \pi G \varrho }{3} = \frac{k}{a^2}+
\frac{\dot{a}^2}{a^2}+ 2 \, \frac{\dot{\lambda} \, \dot{a}}{\lambda \, a}+
\frac{\dot{\lambda}^2}{\lambda^2} - \frac {\Lambda_{\mathrm{E}} \lambda^2}{3} \,, \label{E1p} \\
-8 \, \pi G p = \frac{k}{a^2}+ 2 \frac{\ddot{a}}{a} + 2 \frac{\ddot{\lambda}}{\lambda}+\frac{\dot{a}^2}{a^2}
+ 4 \frac{\dot{a} \, \dot{\lambda}}{a \, \lambda}-\frac{\dot{\lambda^2}}{\lambda^2} -\Lambda_{\mathrm{E}} \,  \lambda^2  \, .
\label{E2p}
\end{eqnarray}
\noindent
These equations contain several additional terms with respect to the standard case, 
these are  those with the scale factor $\lambda(t)$  and its time derivatives.   
The equations also contain the Einstein cosmological constant $\Lambda_{\mathrm{E}}$, 
which corresponds to  the energy density of the vacuum \citep{Carroll92}, see also Appendix B.
In these equations, as well as in Eq. (\ref{field}), $\Lambda_{\mathrm{E}}$ 
is multiplied by $\lambda^2$, the product of the  two represents  the cosmological 
constant $\Lambda$ in the scale invariant space \citet{Canuto77}.
A new  interpretation of the cosmological constant problem has been proposed
within the multiverse approach of Quantum Cosmology. It reconciles the Planck-scale
huge vacuum energy–density predicted by quantum physics
with the observed small value of $\Lambda_{\mathrm{E}}$ \citep{GueorM20}.

The field equation (\ref{field}) and cosmological equations (\ref{E2p})
are undetermined due to their  gauge symmetry. The same problem appears in GR where the undeterminacy is resolved by the choice of 
a line element such as de Sitter, FLWR, etc, which defines the geometry of the system to study. Here, we have to choose some gauging condition
to fix the scale factor $\lambda$. \citet{Dirac73} and \citet{Canuto77} were chosing the so-called ``Large Number Hypothesis'' 
to fix the gauge. As the cosmological solutions depend on the choice of the gauge, this choice plays a key role, to some extent 
as the choice of the metric  in GR. Here, in the scale
invariant framework, both the metric and the gauging condition 
play a major role.

We adopt as  basic gauging  condition the
 assumption of  the scale invariance of the empty space. We notice
 that this choice is consistent  with the remarks by \citet{Feynman63},
 that the occurrence of  scale references are closely related to the matter content of the system. 
 Thus,  the hypothesis of the scale invariance of the empty space (at macroscopic scale) is quite justified.
 Also, as shown by \citet{Carroll92}
the usual equation of state for the vacuum $p_{\mathrm{vac}}= -\varrho_{\mathrm{vac}}  c^2$  is precisely 
the relationship permitting  the vaccuum density   to remain constant for an adiabatic
expansion or contraction.  Thus, it is not so surprising 
that this gauging conditions leads to an analytical relation between
the scale factor $\lambda$ and 
the cosmological constant, which represents the energy density of the vaccum.  

Imposing scale invariance of the empty space means to set the Ricci tensor ($R'_{\mu\nu}=0$) 
and the energy-momentum tensor for matter ($T_{\mu\nu}=0$) all to zero.
By considering the surviving non-zero time and space components \citep{Maeder17a,MaedGueor20a}
one arrives at the specific relations between the scale factor
and the Cosmological Constant  $\Lambda_{\mathrm{E}}$:
\begin{eqnarray}
\  3 \, \frac{ \dot{\lambda}^2}{\lambda^2} \, =\, \lambda^2 \,\Lambda_{\mathrm{E}}  \,  
\quad \mathrm{and} \quad  \,  2\frac{\ddot{\lambda}}{\lambda} - \frac{ \dot{\lambda}^2}{\lambda^2} \, =
\, \lambda^2 \,\Lambda_{\mathrm{E}}\,, \label{diff1}\\
\mathrm{or} \quad \frac{\ddot{\lambda}}{\lambda} \, = \,  2 \, \frac{ \dot{\lambda}^2}{\lambda^2} \, , \quad
\quad \mathrm{and} \quad \frac{\ddot{\lambda}}{\lambda} -\frac{ \dot{\lambda}^2}{\lambda^2} \,
= \, \frac{\lambda^2 \,\Lambda_{\mathrm{E}}}{3} \, .
\label{diff2}
\end{eqnarray}
\noindent
The second group is a variant formulation obtained from the first one.
The two equations ({\ref{diff1}) are just what remains from the time and space components 
of the general scale invariant field equation (\ref{field}) expressed for the empty space.
It is also worth to emphasize  that these equations 
give a new significance to the cosmological constant. In particular, the cosmological constant appears 
as defined by the relative variations of the scale factor (cf. the first one in Eqs.(\ref{diff1}).

As discussed in \citep{Maeder17a} the general solution of (\ref{diff2}) is of the form
$\lambda(t)= a(t - b)^n + d$ and resolves in $d=0$, and $n=-1$ which by setting $t=t_0$ 
results in $a=(t_0-b)\lambda_0$. If the units are chosen so that $\lambda_0=1$ then
this provides interpretation of $b$ as the moment in the past where $\lambda$ blows to 
infinity and $a$ is then the time since then. By choosing this special moment to be at $t=0$
we can set $b=0$ and thus obtaining $\lambda(t)=\lambda_0 (t_0/t)$ as specific choice of
time keeping. Thus, these equations impose a variation of the $\lambda(t)$-term like $t^{-1}$, 
and we choose a normalization constant so that $\lambda= 1$ at the present time $t_0$.
 Thus, we have
\begin{equation}
\lambda(t) \, = \,  \frac{t_0}{t} \, .
\label{lt}
\end{equation}
 In the Appendix A, we further comment on this form of the scale factor, showing it is consistent with 
the most general possible expression for it.

Interestingly enough, Eqs. (\ref{diff1}) and (\ref{diff2}) lead to noticeable  simplifications of  equations (\ref{E2p}):
\begin{eqnarray}
\frac{8 \, \pi G \varrho }{3} = \frac{k}{a^2}+\frac{\dot{a}^2}{a^2}+ 2 \,\frac{\dot{a} \dot{\lambda}}{a \lambda} \, ,
\label{E1} \\
-8 \, \pi G p  = \frac{k}{a^2}+ 2 \frac{\ddot{a}}{a}+\frac{\dot{a^2}}{a^2}
+ 4 \frac{\dot{a} \dot{\lambda}}{a \lambda}  \, .
\label{E2}
\end{eqnarray}
\noindent
 A  third equation may be derived from the above two,  
 \noindent
\begin{equation}
- \frac{4\pi G}{3} \left(3p+\varrho \right) = \frac{\ddot{a}}{a} + \frac{\dot{a} \dot{\lambda}}{a \lambda} \, .
\label{E3n}
\end{equation}
Since $\dot{\lambda}/{ \lambda}$ is negative,  the extra term leads to a repulsive force also depending on the expansion rate.
This is the force responsible for  the acceleration of the expansion of the
model Universe illustrated in Figure \ref{rates}.

The cosmological constant has now disappeared from the equations. In Eq.(\ref{E1p}), the last two terms have cancelled,
and in Eq.(\ref{E2p}) the third and last two terms  have done the same. The consequence is that the product $\dot{\lambda}/\lambda$
now only appears multiplied by the Hubble expansion rate $\dot{a}/a$. 
If the factor $\lambda$ is constant, one is brought back to the  Friedman equations. 
We will see from the solutions in Fig. \ref{rates} that, quite consistently,  the range of variation of this term is larger for the
lower density models and vanishes for models with density tending towards the critical one.
The properties of these equations as well as their solutions have been
discussed  before \citep{Maeder17a}. 

In the case of energy-density dominated by radiation and relativistic matter, for flat scale invariant models with $k=0$,
analytical solutions  for the expansion factor,
the matter density, radiation density and temperature have been obtained by \citet{Maeder19}.
The interesting point is that in the early phases, the main functions 
have the usual dependence on the age $\tau=t-t_\mathrm{in}$ of the Universe as measured from the 
initial time of the Universe at the Big Bang $t_\mathrm{in}$ when the scale factor is zero $a(t_{\mathrm{in}}) = 0$.
For example, the expansion factor goes like $a(\tau) \sim \tau^{(1/2)}$
and then the temperature like $T  \sim  \tau^{-(1/2)}$. 
The run of other variables and the numerical coefficients of the different  
analytical relations near the origin are given for density parameters $\Omega_{\mathrm{m}}$ 
between 0.04 and 0.50 \citet{Maeder19}.

\begin{figure*}
\centering
\includegraphics[width=12.5cm, height=8.5cm]{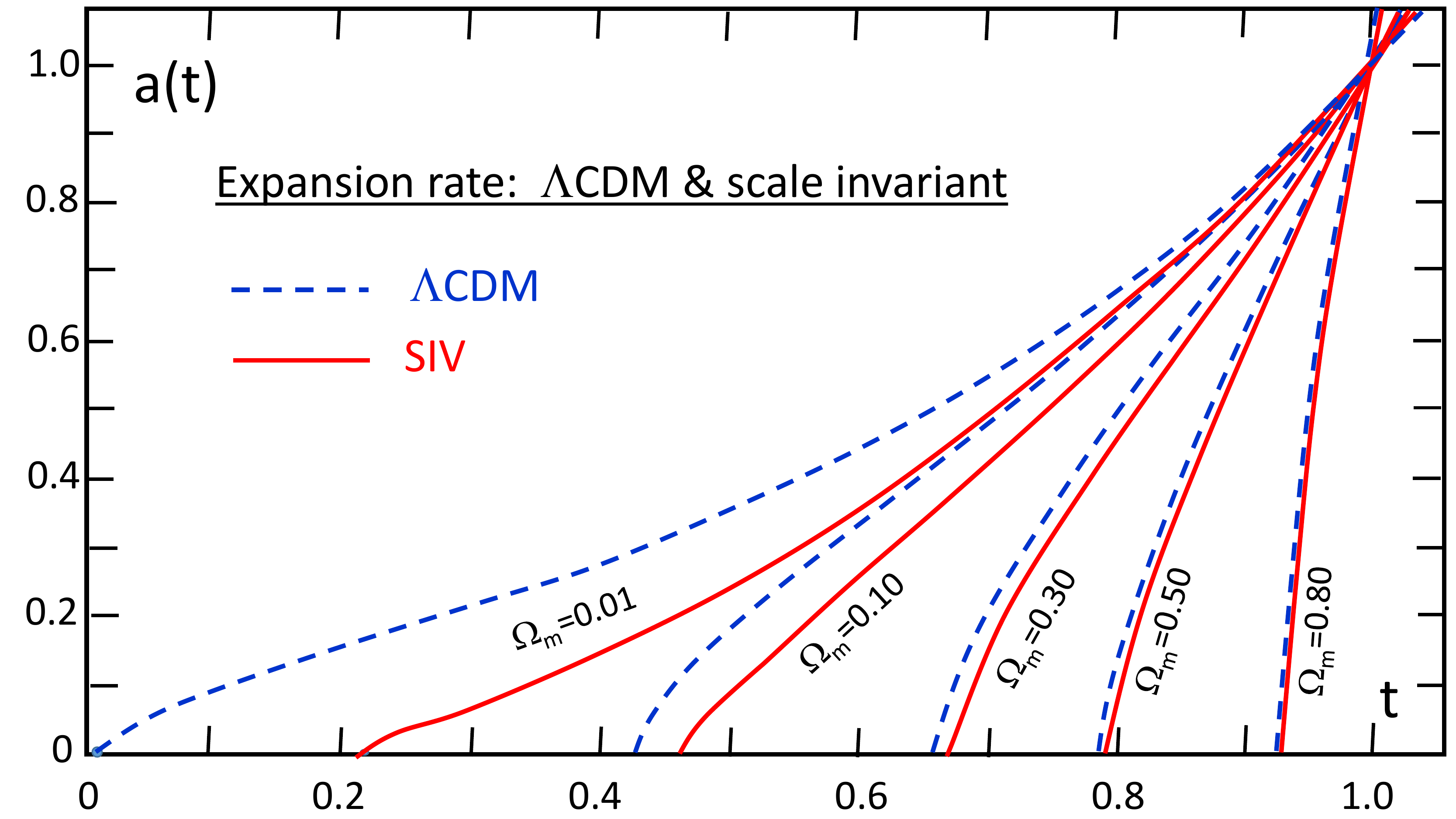}
\caption{The expansion rates $a(t)$ as a function of time $t$ for the flat ($k=0$) $\Lambda$CDM  and SIV 
models for the dominated matter era. 
The curves are labeled by the values of $\Omega_{\mathrm{m}}$.  We notice that the differences between the two sets 
of models rapidly declime for increasing matter densities. }
\label{rates}
\end{figure*}

In the case of flat scale invariant matter dominated models with $k=0$,
analytical solutions have been obtained by \citet{Jesus18},
\begin{equation}
a(t) \, = \, \left[\frac{t^3 -\Omega_{\mathrm{m}}}{1 - \Omega_{\mathrm{m}}} \right]^{2/3}\, ,
\label{R}
\end{equation}
\noindent
in agreement with numerical solutions  \citep{Maeder17a}.
At  present, the time  $t_0$ is fixed   to $t_0=1$ and the expansion factor to $a(t_0)=1$.

The density parameter $\Omega_{\mathrm{m}}$ and the critical density $\varrho_{\mathrm{c }}$ 
are defined according to the usual expressions,
\begin{equation}
\Omega_{\mathrm{m}}= \frac{\varrho}{\varrho_{\mathrm{c }}}\, \quad  
\mathrm{with} \quad  \varrho_{\mathrm{c }}= \frac{ 3 \, H_0^2}{8 \pi G} \, .
\label{rhocc}
\end{equation}
\noindent
$\Omega_{\mathrm{m}}$ vary with time. 
It is  generally considered at the present epoch and so is $H_0=H(t_0)=\dot{a}/a$.
We also write  
\begin{equation}
\Omega_{\mathrm{k}} = -\frac{k}{a^2  H_0^2} \, \quad \mathrm{and} 
\quad \Omega_{\lambda} \, =\,- \frac{2}{H_0} \left(\frac{\dot{\lambda}}{\lambda}\right)_0 \,= \,\frac{2}{ H_0 \, t_0} \,,
\label{hl}
\end{equation}
\noindent
With these definitions, the cosmological equation (\ref{E1}) leads to 
an expression that is valid at all times, since the beginning of time
$t=0$, and in particular at the current time as well ($t=t_0$):
\begin {equation}
\Omega_{\mathrm{m}} \, + \, \Omega_{\mathrm{k}} \, +  \Omega_{\lambda} = \, 1  \, .
\label{Omegapr}
\end{equation}
\noindent
The terms respectively are the contributions to the total energy density 
of the matter, space curvature, and scale factor $\lambda$ energy densities
as normalized to the critical density $\varrho_{\mathrm{c }}$. 
In agreement with the usual practice, the $\Omega$-parameters of the models (for example in Eq. (\ref{R}) generally 
 represent the values at the present time $t_0$. Thus,
to avoid unnecessary index clutter, we will not be using a sub-index $0$ for the present values
of these contributions, except for the traditional use of such index in $H_0$, 
in favor of carefully paying attention to the time moment to be considered.
In a similar manner, we will usually use units such that $t_0=1$, 
unless we feel that the use of explicit $t_0$ in the formulas is more appropriate.

\subsection{Constraints on the mean model density} \label{density}
The scale invariant models are themselves placing constraints on the validity of the scale invariance hypothesis.
Following the Boomerang experiment  \citep{deBern00}, we are considering flat Universe models
with $\Omega_{\mathrm{k}}=0$. From Eq. (\ref{Omegapr}), we  have then,
\begin {equation}
\Omega_{\mathrm{m}} \, +  \,  \Omega_{\lambda} = \, 1  \,.
\label{Omegapr2}
\end{equation}
\noindent
Let us consider a sequence of models of increasing  matter density with
$\Omega_{\mathrm{m}}$ values from 0 to 1.0. 
For $\Omega_{\mathrm{m}}=0$,  the cosmological               
model shows  a maximum relative variations of $( {\dot{\lambda}}/{\lambda})$, corresponding to $\Omega_{\lambda}=1$.

For increasing $\Omega_{\mathrm{m}}$, there is a range of decreasing values of $\Omega_{\lambda}$, and thus 
the relative variations of $\lambda$ are more and more  limited according
to the above equation  (\ref{Omegapr2}).
In this domain, there are possible solutions of the scale invariant equations  (\ref{E1})  and (\ref{E2}).
For  the particular case $\Omega_{\mathrm{m}}=1.0$ (at present), 
Eq.(\ref{Omegapr2}) implies that  $\Omega_{\lambda}$  is zero,
which means that $\lambda$ is a constant (at  present, but also at all times according to
form of $\lambda$). 
Thus, we see that Eqs. (\ref{E1}) and  Eqs. (\ref{E2})  become identical to the Friedman equations, 
and  in  this case we would get the same solution for the scale factor $a(t)$ as function of $t$,
that is,  $a(t)\propto\;t^{2/3}$  and $a(t)\propto\;t^{1/2}$ for the matter and radiation eras respectively.

For $\Omega_{\mathrm{m}}  \geq 1$,the situation is completely different, 
one would have $\Omega_{\lambda} <0$,  which 
according to Eq. (\ref{hl}) leads to,
\begin{equation}
\frac{2}{H_0 \, t_0} \, < 0 \,.
\label{hneg}
\end{equation}
\noindent
Such a condition would imply  either  a contracting Universe or a negative time, which although formally
not impossible, do not correspond to  the observed  Universe.
Thus,  we conclude that scale invariance is problematic  for a  mean density 
larger than the critical density $\varrho_{\mathrm{c}}$,
(unless we are willing to consider local contraction with $\kappa_0=-\dot{\lambda}/\lambda<0$
as a process of transfer of energy into the mechanical degrees of freedom related to the gravitational interactions, 
such us black hole formation).

Let us make a brief comment on the initial time of the Big-Bang.
It is    given by $a(t_{\mathrm{in}}) = 0$ is,
\begin{equation}
t_{\mathrm{in}} \, = \, t_0 \, \Omega^{1/3}_{\mathrm{m}} \, .
\label{tin}
\end{equation}
Where, we explicitly use $t_0$  (usually equal to 1) in these expressions to indicate the correctness of the units.
In the  radiative era, the behaviour of $a(\tau)$  close to the origin is in $\tau^{1/2}$, where $\tau$ is the age.
 In a scale where $t_0=1$,
the differences between $t_{\mathrm{in}}$ given by Eq.(\ref{tin}) and by a detailed  modeling of the radiative era is very small.
For example, for $\Omega_{\mathrm{m}}=0.20$, Eq.(\ref{tin}) gives $t_{\mathrm{in}}=0.5848036$, 
while the complete model with the radiative era gives $t_{\mathrm{in}}=0.5848043$ \citep{Maeder19}. 

The reason for the very small difference is that the transition time  $t_{\mathrm{eq}}$  
from the matter to the radiation era occurs very closely to $t_{\mathrm{in}}$.
In the present example, it occurs at $t_{\mathrm{eq}}= 0.5848066$.
Thus, the  difference of slopes in the time interval ($t_{\mathrm{eq}} - t_{\mathrm{in}}$) 
only leads to a very tiny difference in the time at which $a(t) = 0$. 
We may note that such properties are typical of current cosmological models.

The Hubble expansion rate  $H(t) = \dot{a}/a$, based on (\ref{R}) is:
\begin{equation}
H(t) \, = \, \frac{2 \, t^2}{t^3 - \Omega_{\mathrm{m}}} \, , \quad \mathrm{with} 
\quad  H_0 \, = \, \frac{2}{1-\Omega_{\mathrm{m}}} \, .
\label{Hh}
\end{equation}
\noindent
From Equation (\ref{Hh}), a value  $\Omega_{\mathrm{m}} \geq 1$ would 
imply a contracting universe since  the Hubble  rate $H_0$ would be negative. This would be  in severe contradiction
with all recent determinations, {\it{e.g}} the SH0ES 
collaboration using the cosmic distance ladder method gives a value $H_0= 73.2\pm1.3\, \rm km/s/Mpc$ \citep{Riess21}.
There would be expanding solutions ($H_0>0$) only for $t < \Omega^{1/3 }_{\mathrm{m}}$, 
{\it{i.e.}} before $t_{\mathrm{in}}$,  the Big-Bang.
Even worse, the initial time would appear equal or larger than  the present time!  
Thus,  there  is clearly no realistic model, of the observed Universe,
with $k=0$ and a density  equal or larger than the critical density $\varrho_{\mathrm{c}}$. 
Thus, the overall conclusion is that 
{\it{for flat models, scale invariance is  possible for   mean densities
below the critical  density $\varrho_{\mathrm{c}}$, and   is forbidden above  $\varrho_{\mathrm{c}}$.}}
\footnote{A recession velocity of $70$ km s$^{-1}\, $ Mpc$^{-1}$ corresponds to 
$2.269 \cdot 10^{-18}\,$ s$^{-1}$, the inverse of an age of  $4.408 \cdot 10^{17}$ s or 13.97 Gyr.
This leads to $\varrho_{\mathrm{c}} = 9.207 \cdot 10^{-30}$  [g cm$^{-3}$].}
This is the answer of the  general field equation, 
however this does not necessarily mean that this is the case in the real Universe.
The definite answer may only come from observations.

Figure  \ref{rates} shows the  expansion rates in the matter dominated era 
for the scale invariant and $\Lambda$CDM model results for $k=0$ 
and  various values of $\Omega_{\mathrm{m}}$ between  0.01 and 0.80.
For $\Omega_{\mathrm{m}}=0$,  the  empty scale invariant model has a growth rate  $ a(t)  \, \sim \,t^2$. 
As seen above, a value   $\Omega_{\mathrm{m}}= 1$ implies $\dot{\lambda}=0$  
and the system of scale invariant equations is brought  back to the  Friedman  system with 
$a(t) \, \sim \,t^{2/3}$ in the matter dominated era.
Both sets of models  containing matter start explosively  near the origin 
with very high values of $H={\dot{a}} /a$ and a positive value of $q=-\ddot{a}a/(\dot{a})^2$, indicating braking.
After the  initial  braking phase,  both sets of  models with $0 < \Omega_{\mathrm{m}} < 1$ show 
an accelerated expansion, which after an inflection point for  $q=0$ goes on all the way. 
The inflection occurs later  for higher densities. In the case of scale invariance with $k=0$, the inflection point occurs at 
a time $t_{\mathrm{q=0}}$ according to Eq. (\ref{R}),
\begin{equation}
t_{\mathrm{q=0}} \, = \, (2 \, \Omega_{\mathrm{m}})^{1/3} \,,
\end{equation}
and with  an expansion factor,
\begin{equation}
a(t_{\mathrm{q=0}}) \, = \, \left( \frac{ \Omega_{\mathrm{m}}}{1- \Omega_{\mathrm{m}}} \right)^{2/3} \,.
\end{equation}
The differences between the SIV and $\Lambda$CDM models in Fig. \ref{rates}
are  rapidly reduced for increasing values of $\Omega_{\mathrm{m}}$ up to 1.0. 
Above a density equal to about 30\% of the critical density, the differences
 become very small. This clearly demonstrates how fast the effects of scale invariance disappear when the matter density
progressively increases at very low densities below $\varrho_{\mathrm{c}}$.

\section{The  horizons and their limitation on scale invariance}  \label{horiz}

In order to proceed further, we need to specify the notions of horizons for the scale-invariant models. 
Since scale invariance is likely prevented by causality relations, 
which permit gravity and light information to be transferred from one place in the Universe to another one. 
Thus, the occurrence or not of scale invariance may be related to the problem of horizons. 
We follow the definitions and properties of the horizons as clarified by \citet{Rindler56,Rindler69}.

\subsection{The case of the particle-horizon }   \label{partic}

At a given age of the Universe, the finite velocity of light limits the distances from where we may receive electromagnetic
signals. For a signal emitted in $r_1$ at time $t_1$ and received in $r=0$ at time $t_0$, one  has the following
relation, obtained from the equation of photon propagation $ds= 0$,
\begin{equation}
\int_0^{r_1} \frac{dr}{\sqrt{1-kr^2}} \, = \, \int_{t_1}^{t_0} \frac{dt}{a(t)}\,.
\label{ph}
\end{equation}
where  the FLWR metric has been used.
If the integral on the right converges when $t_1$  tends to $0$, {\it{i.e.}} at the origin, then
the left integral also converges  and  $r_1$ tends towards a finite value $r_{\mathrm{H}}$,
called the \textit{particle-horizon} \citep{Rindler69}. Thus, at a given time the limit $r_{\mathrm{H}}$ separates the Universe
in two parts, one from which we may receive information and one inaccessible. On the contrary, if the right integral diverges,
$r_{\mathrm{H}}$ tends towards infinity and signals may theoretically be received, at present, 
from the whole Universe. As an example, for a law 
$a(t) \, \sim \, t^n$, if $n \, < \, 1$, the integral on the right  is converging. 
As an example, the Einstein-de Sitter (EdS) model with $n=2/3$ is converging (the same for $n=1/2$). 
It has a particle horizon at a distance,
\begin{equation}
d_{\mathrm{H}}(t_0)\, = \, 3 \, c\, t_0 \, .
\label{tct}
\end{equation}
For a static Universe, the distance would just be $c \, t$, but due to the fast early expansion this distance is larger.
The EdS model has a particle-horizon,  meaning that at a time $t$ the light of objects more distant than $3 \, c \,t$ 
has not yet reached us.  The expansion rate $ a(t) \sim t^{2/3}$ ``progresses'' slower than the horizon which does it  like $ t$.
It implies that as time goes more distant objects enter our horizon.\\

The expansion factor $a(t)$ for scale invariant models is given by Eq. (\ref{R}).
Let us examine the situation  for $0 < \Omega_{\mathrm{m}} <  1.0$.
It is sufficient to study the behavior  of Eq. (\ref{R})  near the origin.
We write  the time near the origin  as $ t \, = \, \Omega_{\mathrm{m}}^{1/3} + \delta t$, 
and get for $\delta t  \rightarrow 0$,
\begin{equation}
\left(t^3- \Omega_{\mathrm{m}}\right)^{2/3} = \left( 
3  \Omega^{2/3}_{\mathrm{m}} \delta t+3 \Omega^{1/3}_{\mathrm{m}} \delta t^2 +\delta t^3 \right)^{2/3} 
\propto  \, \left(3 \Omega^{2/3}_{\mathrm{m}} \delta t \right)^{2/3}.
\end{equation}
Thus, the exponent  $n$ in $a(t) \sim t^n$ near the origin, but still in the matter dominated era, is $n=2/3$,
there the scale invariant model behaves like the EdS model. In the example given in Sect. \ref{density}
for $\Omega_{\mathrm{m}}=0.20$, 
the transition to the radiation era occurs at  a redshift $z=4028$.
In  the radiation era,  the behaviour of $a(t)$ is  in $t^{1/2}$,  the conclusion is the same in both cases.
The integral is converging, implying that  the scale invariant models  with $\varrho \, < \varrho_{\mathrm{c}}$ 
have a particle-horizon.  

\subsection{The case of the event-horizon}   \label{event}

The existence of a particle-horizon implies that there are domains of the Universe which cannot be observed now.
However, some domains may progressively become accessible as time is going. 
There could also be domains which will never become 
accessible, even after an infinite time. Let us again consider Equation (\ref{ph}).  Now, we consider the case 
where $t_0$ tends towards infinity. If the integral on the right is converging,
there is a limit $r_{\mathrm{E}}$ in the left integral, called the \textit{event-horizon} \citep{Rindler69},
beyond which  the events will never reach us. This occurs for $n > 1$  for an expansion rate $a(t) \, \sim \, t^n$.
This means that the  expansion is accelerating  and, thus that 
some regions of the Universe presently visible (at least in theory)
are progressively  getting  out of accessibility.
On the contrary, if the integral on the right is diverging, 
$r_{\mathrm{E}}$ tends towards infinity and  the concerned models have no event-horizon: 
all the domains of the Universe will become accessible in the future.
This occurs if $n < 1$ , the horizon ``advances faster'' than the expansion.

The scale invariant models with $0 < \Omega_{\mathrm{m}} <  1.0$, after an initial braking phase,
experience an  acceleration, alike the $\Lambda$CDM models.
As matter, and also radiation as well, become diluted these models are 
progressively tending towards a behavior in  $t^2$ and $e^{Ht}$ respectively and 
thus have an event-horizon.

In summary, the scale invariant models with $0 < \Omega_{\mathrm{m}} <  1.0$, 
have near the origin a behavior  alike the EdS model  (in $t^{2/3}$ and $t^{1/2}$) and thus    have a particle-horizon. 
For large enough times, after an inflection point (which depends  on $\Omega_{\mathrm{m}}$), 
they are accelerating and thus also have an event-horizon.
Matter is entering the particle-horizon in the early phases and  
getting out in the later phases.  Thus, these Universe models, on both sides of the arrow of time, have regions which are
not causally connected. Apart from the question of inflation (Sect. \ref{infl}), this  lets open the door for scale invariance.

\subsection{Physical conditions for scale invariance}   \label{phys}

We   have to understand why the critical density $\varrho_{\mathrm{c}}$ appears as a limit
above which the effects of scale invariance are absent from cosmological models. Below $\varrho_{\mathrm{c}}$,
the equations are permitting scale invariance. It is an interesting  possibility, but not a proof of existence.

Let us consider an observer in an homogeneous  and isotropic  medium of mean density $\varrho$,
expanding according to the Hubble-Lemaître law with a  present expansion rate $H_0$. 
At  some limiting distance $R_{\mathrm{lim}}$ from the observer,
the recession velocity  becomes equal to  the light velocity,
\begin{equation}
H_0 \, R_{\mathrm{lim}}  \, \simeq \, c \, .
\label{rc}
\end{equation}
In the relativistic context, a recession velocity equal to $c$ corresponds to an infinite
redshift. $ R_{\mathrm{lim}}$ is also a meaningful definition for a horizon, it may be called the ``\textit{physical horizon}''.
No gravity effect, no gravitational or electromagnetic waves from larger distances can reach the observer.

We note that  $R_{\mathrm{lim}}$ differs  from the formal definition of the particle-horizon 
given in Sect. \ref{partic}. For example, in the EdS model,
one has with $H_0= (2/3) (1/t_0)$,
\begin{equation}
\frac{2}{3 \, t_0} \, R_{\mathrm{lim}} \, = \, \frac{d_{\mathrm{H}}}{3 \, t_0}\, , \quad \quad \mathrm{thus}  \; \quad
\ R_{\mathrm{lim}}\, = \frac{1}{2} \, d_{\mathrm{H}}\,.
\end{equation}
In general, $R_{\mathrm{lim}}$ is  smaller  than $d_{\mathrm{H}}$. This is due to the fact 
that the particle-horizon formally goes  back to time zero, 
where  the initial    expansion rate is extreme,  tending towards infinity at the initial singularity (even without inflation).
This is the case in the EdS model, as well as in the  scale invariant models with $0 < \Omega_{\mathrm{m}} <  1.0$. 
Both the particle-horizon and event-horizon  depend on model properties  and in particular on those of the most extreme
phases, the initial one for the particle-horizon and the final one for the event-horizon.
The physical horizon $ R_{\mathrm{lim}}$ has the advantage of not resting on a particular model of the
extreme phases. It is  model independent, apart from the fact that the Hubble law 
assumes an isotropic and homogeneous Universe in expansion. 
Thus,  $R_{\mathrm{lim}}$ may be considered 
as the meaningful, model independent, horizon for causality connection. 

Over distances smaller than  the physical horizon, causality connection is present, 
gravity effects are acting, electromagnetic waves are  transmitted, etc. 
This means that for distances smaller than $R_{\mathrm{lim}}$,
scale invariance is likely forbidden since physical connection is present over the whole domain within this limit.
For distances larger  than $R_{\mathrm{lim}}$,  the physical connection is absent
and  scale invariance might be present.
With the expression of the critical density $\varrho_{\mathrm{c}}$ given in Eq. (\ref{rhocc}),
the above equation (\ref{rc}) becomes,
\begin{equation}
\frac{8 \pi}{3} \,G \,\varrho_{\mathrm{c}} \,  R^2_{\mathrm{lim}} \, \simeq \, c^2 \, .
\label{rs}
\end{equation}
If $\varrho_{\mathrm{c}}$ is the mean density of the matter within $R_{\mathrm{lim}}$,
{\it{i.e.}} if  $\varrho_{\mathrm{c}}= \frac{3 M}{4 \pi \, R^3_{\mathrm{lim}}}$,  the limiting distance $R_{\mathrm{lim}}$
is equal to the Schwarzschild radius $R_{\mathrm{S}} = 2\,GM/c^2$.
This would correspond to $\Omega_m=1$ and it is also the 
highest possible matter density for an object in our Universe - a black hole.
This can be seen even from classical Newtonian considerations presented by \cite{Freeman'75}.
Equation (\ref{rs})  leads to the following value of the limit radius in terms of the critical density,
\begin{equation}
R_{\mathrm{lim}} \, \simeq \, c \, \sqrt{\frac{3}{8 \pi \,G\, \varrho_{\mathrm{c}}}} \, 
\label{rlim}
\end{equation}
If the real mean density $\varrho$ of the medium is higher than the critical density $\varrho_{\mathrm{c}}$,
the same amount of mass  (whatever it is)  has a distribution  in space, which is
contained in a radius $R$ smaller than $R_{\mathrm{lim}}$. 
Thus, the whole volume enclosed  within radius $R$ is causally connected. 
Thus, physical  units  are defined throughout.  The above simple model 
suggests  that a medium with  $\varrho \, \geq \, \varrho_{\mathrm{c}}$ is unlikely to be   scale invariant,
since its various parts are physically connected, being enclosed within $R_{\mathrm{lim}}$.
This throws some light on  the above results from the  scale invariant  equations, which showed
the absence of SIV models with $\Omega_{\mathrm{m}} > 1$,
since such situation will imply an object denser than a black hole.
Thus, according to (\ref{Omegapr2}) and (\ref{hl}) the conformal factor $\lambda$ would be increasing,
which is the reversal process of the usual behavior of  $\lambda$ as seen in (\ref{lt}).

At the opposite,  when  $\varrho \,  < \, \varrho_{\mathrm{c}}$,
the considered  volume  cannot lie entirely within the limit  $R_{\mathrm{lim}}$. This means that,
while some parts are  connected, causal connection by gravity and light is not present in 
the whole system.  Moreover, as such systems are accelerating,  
some domains of the accessible Universe will escape in future. 
This lets open the possibility that space-time is   scale invariant for $\Omega_{\mathrm{m}} < 1$,
a fact that would be in agreement  with the existence of solutions to the 
equations  (\ref{E1}) and (\ref{E2}), but  as repeatedly mentioned  this  is not a proof.

The problem is also related to the cosmological history of the Universe.
In this respect, near the origin, the expansion rate $H$ of the   models, whether $\Lambda$CDM, EdS or scale invariant,
are diverging.  with an expansion  faster than that of $R_{\mathrm{lim}}$. This would  favor matter outside the horizon.
Such matter outside the horizon is consistent with the black-hole universe idea and 
may not possess many of the problems of the Standard Big-Bang (SBB) model 
while not necessarily requiring a long period of cosmological inflation \citep{EassonBrandenberger01}, 
and even be also consistent with the multiverse ideas where
global structure of the spacetime contains an infinite sequence of black and white holes, 
vacuum regular cores and asymptotically flat universes \citep{Dymnikova et al. 2001}.
The region around the singularity at the center of a black hole would naturally provide confined high-energy density 
and therefore the needed high potential energy for inflation. It has been argued that the coupling between 
the spin of elementary particles and torsion in the Einstein-Cartan theory of gravity 
generates gravitational repulsion at extremely high densities in fermionic matter, 
approximated as a spin fluid, and thus avoids the formation of singularities in black holes.
It may even undergo  several nonsingular bounces until it has enough matter to reach a size at which 
the cosmological constant starts cosmic acceleration with 
a finite period of exponential expansion (inflation) of such universe creation \citep{Poplawski16}.
There are many such interesting open  questions regarding the singularity.  They  are however
beyond the scope of the present work and we now concentrate on the 
 the main question concerning the inflation, in particular whether  this phase of incredibly fast explosion 
is compatible  and predicted by scale invariant equations. This critical question  is examined next.

\section{Inflation, conservation law and scale invariance} \label{infl}

The high isotropy of the CMB radiation has brought a problem for the various current models with  a particle-horizon. 
Regions separated on the sky by an angle larger
than about two  degrees were outside their own horizons at $z \sim 1100$ on the last scattering surface.
Thus, the very high isotropy, that the whole CMB sky is presenting, was difficult to explain.
The inflation theory \citep{Guth81}  predicts an initial exponential  growth of the initial Universe at  Planck length
by a factor $e^N$, with typically $N > 62$ during the first $10^{-32}$ s or so,
see reviews by \citet{Linde95,Linde96,Linde05}, \citet{Weinberg08}. Thus, physical
interactions between the different parts  of the Universe were possible before the inflation. The inflation also accounts  for
the observed flatness of the Universe (and  for the  incredibly higher
accuracy with which it had to be satisfied  in  early stages).
Also, the absence of magnetic  monopoles  is considered as a consequence of the inflation. In addition, 
the inflationary Universe is  the source
of the spectrum of primordial fluctuations  \citep{Kofman88}, see also review by  \citet{Coles02}.
We first examine  the  relations between  the scalar field associated to  the scale invariance  and  the scalar field of the inflation, 
and  then whether  the inflation is compatible and  may also occur in the context of the scale invariant equations.

\subsection{The energy density of the vacuum and the cosmological constant}

Let  $\ell'$  be some  constant line element in the space of GR. 
In the scale invariant space,  the corresponding line element $\ell$  
behaves as $\ell \, = \ell'/\lambda(t)$,  
where the scale factor $\lambda$ is only a function of the cosmic time $t$, as said above.  
The possible variations of the scale factor $\lambda(t)$ may  contribute to
the  energy density present in the empty space. 
If  $\lambda(t)$ varies,  the  energy  associated  to the length $\ell$ 
in the  empty space will be given by an expression related to its change:
\begin{equation} 
\ell \, = \frac{\ell'}{\lambda}\;\Rightarrow\; 
\dot{\ell}^2 \,  = \ell^2  \frac{\dot{\lambda}^2}{\lambda^2}\, .
\label{l}
\end{equation}
The energy density $\varrho$ in the scale invariant space is obtained by taking the above value by length unit.
Thus, if there is no other source of energy in the  empty space, 
 its energy density $\varrho$ can be written
\begin{equation}
\varrho \,  \sim \, \frac{1}{2} \, \frac{ \dot{\ell}^2}{\ell^2}  \, \quad   \mathrm{and \; thus} \quad \varrho= 
\,  \frac{1}{2} \,  C\,   \frac{\dot{\lambda}^2}{\lambda^2}  \, ,
\label{ro} 
\end{equation}
where  $C$ is a proportionality constant, which has to be fixed in a consistent way with current definitions.

The Einstein cosmological constant $\Lambda_{\mathrm{E}}$ 
is related  to the energy density $\varrho'$ of the empty space in GR  \citep{Carroll92},
\begin{equation}
\Lambda_{\mathrm{E}} \,= \,  8\,  \pi \,G \,\,\varrho' \,,
\label{defr}
\end{equation}
In the scale invariant system of Weyl's Geometry, as shown by the field equation (\ref{field}) 
and the cosmological equations (\ref{E1p}) and (\ref{E2p}),
the corresponding cosmological constant $\Lambda$ is,
\begin{equation}
 \Lambda \, = \, \lambda^2 \, \Lambda_{\mathrm{E}}\, .
 \label{Lam}
 \end{equation}
 This is in agreement   with the behavior of the coscalar expressing the relation between
 the vacuum density $\varrho$ in the Weyl's space (noted without prime)  and 
  $\varrho'$ in the Rieman space (noted with a prime)  \citep{Maeder17a},
 \begin{equation}
 \varrho\, = \,   \lambda^2 \, \varrho'\, ,
  \label{varrho}
 \end{equation}
From Eqs. (\ref{diff1}),  $\Lambda_{\mathrm{E}}$ is related to $\lambda$
and its derivatives by,
\begin{equation}
\Lambda_{\mathrm{E}} \,  =\,3 \, \frac{\dot{\lambda}^2}{\lambda^4} \,.
\label{le}
\end{equation}
From Eqs. (\ref{defr}), (\ref{varrho}) and (\ref{le}),  we get  the consistent expression 
of the vacuum density in the scale invariant system,
\begin{equation}
\varrho\, = \,   \frac{3}{8\, \pi \, G} \, \frac{\dot{\lambda}^2}{\lambda^2} \, .
  \label{rhsiv}
 \end{equation}
 Thus, we see that the two  expressions  (\ref{ro}) and (\ref{rhsiv}) of the density of the empty space are consistent  if
 \begin{equation}
 C \, = \, \frac{3}{4\, \pi \, G} \, 
 \label{C}
 \end{equation}
 in the above expression (\ref{ro}).
 In Appendix B,  we further comment on the expression (\ref{rhsiv}) of the vacuum density
 in relation wth the scale invariant cosmological equations.
  
As a side remark, we note that the constancy of 
$\Lambda_{\mathrm{E}}$ also implies Eqs. (\ref{diff2}):
\begin{equation}
\frac{d \Lambda_{\mathrm{E}}}{dt} \,  \sim  \,
2\frac {\dot{\lambda} \ddot{\lambda}}{\lambda^4} 
-4\frac{\dot{\lambda}^3}{\lambda^5}\, = \, 0 \,
\Rightarrow
\frac{\ddot{\lambda}}{\lambda} \, = \,  2 \, \frac{ \dot{\lambda}^2}{\lambda^2} \, .
\label{d1}
\end{equation}
This is the first of  equations (\ref{diff2}), and this shows that the 
present definition of the vacuum density is consistent with the results of the field equations for the empty space.

Both  $\varrho$ and $ \Lambda$ (in the scale invariant space) behave
like $1/t^2$ according to expression (\ref{lt}) based on the field equation of the vacuum. 
This implies  that the energy density of the vacuum, and the cosmological constant $\Lambda$,  {black}
in the scale invariant space become very large near the origin. 
For example at the Planck time $t_{\mathrm{Pl}} = 5.39 \cdot 10^{-44}$ \, s, 
dominated by quantum effects, the cosmological constant would be a factor  
$\left (\frac{4.355 \cdot 10^{17}}{ 5.39 \cdot 10^{-44}}\right)^2 = \, 
6.4 \cdot 10^{121}$ larger than the  value at the present  cosmic age $t = 13.7 \; \mathrm{Gyr} = 4.323 \cdot 10^{17}$ s.
Thus, as such this may solve the so-called cosmological problem by viewing the 
Planck-seed universes and the {$a$-derivable} universes as different stages of the same Universe
rather than a disconnected universe \citep{GueorM20}.
In other words, the smallness of the Einstein cosmological constant $\Lambda_E$  
is naturally related to current age of the Universe, assuming that now $\lambda=1$ by choice of units,
because the solution (\ref{lt}) for (\ref{diff1}) implies 
$\Lambda_E =3/t_0^2\approx 1.6\times10^{-35}\mathrm{s}^{-2}$.

\subsection{The scalar field clock associated to the  scale invariant empty space}   \label{scal}

Instead of the above particular notations, it is appropriate to express the energy-density of the empty
space in term of a scalar field $\psi$, 
\begin{equation}
\varrho \, = \, \frac{1}{2}\, C \,  \dot{\psi}^2  \, \quad \mathrm{with}  
\; \,\dot{\psi}  \, = \,- \frac{\dot{\lambda}}{\lambda} \, .
\label{ro2}
\end{equation}
with the constant $C$ given by Eq.(\ref{C}).
This relation expresses that  the empty space is endowed with an energy density related to scale transformations.
Relation (\ref{lt}) implies,
\begin{equation}
\dot{\psi} = \frac{1}{t} \, . 
\end{equation}  
In other conditions, like at inflation, there could be some 
additional  contribution to the energy-density of the medium.
We note that  $\dot{\psi}$
is in fact equal to the time component of the metrical connection $\kappa_{\nu}$  defined by Eq. (\ref{ka}) and
present in the general scale invariant field equation (\ref{field}),
\begin{equation}
\dot{\psi}  \, = \, \kappa_0   \, , \quad  \mathrm{whith}\quad   \kappa_0= - \dot{\lambda}(t)/\lambda(t) \, .
\end{equation}
The scalar field $\psi$ is thus 
\begin{equation}
\psi \, \sim \, \ln t \,,     \quad \mathrm{and}   \; \;  t \, \sim \, e^{\psi}
\label{psit}
\end{equation}
Near the origin $t \longrightarrow 0$, $\psi \rightarrow  -\infty$ and $\dot{\psi} \rightarrow \infty  $.
Thus, near the origin the time grows very fast as a function of  the field $\psi$ which evolves at a much slower pace
and may thus be the appropriate  ``timekeeping field" near the origin as discussed by \cite{Weinberg89}.
We see below that  field $\psi$ can  be identified with the ``inflaton'',  the rolling field during inflation.

\subsection{The inflation } \label{inflsiv}

Following the first version of inflation by \citet{Guth81}, there were several inflation
theories resting on different hypotheses \citep{Linde95,Linde96,Linde05}, 
{\it{e.g.}}  phase transition in the vacuum,  breaking of grand unified theory, 
specific conditions producing eternal inflation and  chaotic inflation, etc. Over the years, it has emerged that  the key 
condition to have inflation, is the presence of a scalar field  $\varphi$  (called the inflaton)
and a potential $V(\varphi)$, which initially contains most 
of the matter-energy of the Universe \citep{Weinberg08}. At the end of the inflation, the decay of  the potential $V(\varphi)$
is leading to the baryogenesis \citep{Linde95}, the further evolution of which will form the present Universe.
The condition for the existence of an inflation is that at early times 
$V(\varphi)$ is  large and   flat. Then, there are various scenarios  \citep{Linde95}. 
Typically, the scalar field ``rolls'' very slowly at first down this potential, so that the Hubble constant decreases only slowly,
and ``the universe experiences a more-or-less exponential inflation before the field changes very much'' \citep{Weinberg08}.
However, as pointed out by \citet{Brandenberger17}, the whole picture in its different expressions are not free from remaining 
problems.

The energy-density $\varrho$ and the pressure $p$ of the vacuum state 
during inflation  are resulting from  the  contributions of a term of  $\dot{\varphi}^2$
due to the scalar field $\varphi$  and of  an additional  potential $V(\varphi)$,  
which is the dominant one \citep{Linde95,Weinberg08},
\begin{equation}
\varrho \, = \,  \frac{1}{2} \dot{\varphi}^2 + V(\varphi)\,, \quad \mathrm{and} \quad p\,= 
\, \frac{1}{2} \dot{\varphi}^2-V(\varphi) \,.
\label{rp}
\end{equation}
\noindent
We propose to  relate the scalar field  $\varphi$  of the inflation (the ``inflaton'') 
to  the  above  field $\psi$ associated with the  scale invariance of the ordinary empty space 
and to examine the consequences of this identification.  We establish the correspondence
$\sqrt{C} \, \psi \iff \varphi$  and thus we write for the energy-density and pressure at inflation,
\begin{equation}
\varrho \, =  \frac{1}{2} \, C \, \left(\dot{\psi}^2 + U(\psi)\right)\,, \quad \mathrm{and} \quad p\,= 
\, \frac{1}{2} \, C \,\left(\dot{\psi}^2 - U(\psi)\right) \,.
\label{rp1}
\end{equation}
\noindent
We also have the correspondence for the potential with  $ U(\psi)\iff (1/C) V(\varphi)$.
Usually, the potential $V(\varphi)$  is a potential of high energy during inflation. It is supposed not to vary too much during
inflation and various forms have been proposed for it. 
This also applies to  the potential $U(\psi)$ which only differs by a (large) constant factor.
The properties of $V(\varphi)$, and thus of $U(\psi)$,  are supposed to lead to
a Hubble-Lemaître constant $H_{\mathrm{infl}}$ that does not vary too much during some interval of $\psi$ 
and thus tends to produce the exponential growth of the Universe, characteristic of the inflation.
We note that  the relation of $\psi$ with the time $t$ would  naturally explain
why the system is  ``rolling'' slowly along  the  flat potential. \\

As emphasized by \citet{Weinberg08}, the large value of the energy-density during inflation
does not necessarily rule out the classical treatment 
of gravitation according to GR. The quantum gravitational effect may be neglected under the assumption that the energy-density
is already much less than the Planck energy. If  the conditions for the applicability of GR are satisfied, as is usually considered, the
conditions for the applicability of Eq. (\ref{field}) are also met. Once more, it doesn't mean that Nature in her wisdom has done it,
but it is a possibility to be explored. The point to be verified now is whether the scale invariant equations and the above identification 
also permits an inflation.

In the scale invariant context, the equation of energy conservation contains an additional term with respect to the standard case
\citep{Maeder17a},
\begin{equation}
\frac{d(\varrho a^3)}{da} + 3 \, p a^2+ (\varrho+3\, p) \frac{a^3}{\lambda} \frac{d \lambda}{da} = 0 \, .
\label{conserv}
\end {equation}
\noindent
We want to write it in terms of the inflation quantities $\psi$ and $U$. 
For this purpose, it is preferable to start from the following form,
\begin{equation}
\dot{\varrho}+ 3 \, \frac{\dot{a}}{a}\, (\varrho+ p) + \frac{\dot{\lambda}}{\lambda}\, (\varrho+ 3p) \, = \, 0 \, .
\label{conserv2}
\end{equation}
There, we see that the quantities $\varrho$ and $p$ only appear linearly, thus the constant $C$ appears as a multiplicative factor
in front of the left member of this equation. This implies that we can ignore it and thus we have,
\begin{equation}
\ddot{\psi} + U'\,  + 3 H_{\mathrm{infl}} \, \dot{\psi} - 2 \, ( \dot{\psi}^2 - U)\,= \, 0 \,,
\label{cons1}
\end{equation}  
which differs from the usual Klein-Gordon equation \citep{Kofman88,Linde95}
by the presence of the last two terms.
The expansion rate during inflation, {\it{i.e.}} when there is a large potential $V(\psi)$ contributing  to the energy density 
of the vacuum, is noted  here by $H_{\mathrm{infl}}$.
Since  $\ddot{\psi} = - \psi^2$, the above equation simplifies to
\begin{equation}
U'\,  + 3 H_{\mathrm{infl}} \, \dot{\psi} - 3 \dot{\psi}^2 + 2 U\,= \, 0 \,.\label{cons3}
\end{equation}
Thus, the expansion rate behaves like,
\begin{equation}
H_{\mathrm{infl}} \ = \, \dot{\psi} - \frac{2\, U}{3\,\dot{ \psi}}- \frac{U'}{3\, \dot{\psi}} \,.
\label{hh}
\end{equation}
The condition to have an inflation with an exponential growth during a very short time
implies  that the relative change $\mid \dot{H}_{\mathrm{infl}}/H_{\mathrm{infl}}\mid$ 
during a time $1/H_{\mathrm{infl}}$ should be much less than unity \citep{Weinberg08}, {\it{i.e.}}
\begin{equation}
\mid \dot{H}_{\mathrm{infl}} \mid\, \ll  H^2_{\mathrm{infl}}\,.
\label{cond}
\end{equation}
The model Universe  will thus follow a de Sitter-like exponential expansion. 

For our model (\ref{hh}), the time derivative of the expansion rate is:
\begin{equation}
\dot{H}_{\mathrm{infl}} \, = \, \ddot{\psi} -\frac{2 U' \dot{\psi}}{3 \dot{\psi}} +\frac{2 U \ddot{\psi}}{3 \dot{\psi}^2}
-\frac{U'' \dot{\psi}}{3 \dot{\psi}}+ \frac{ U' \ddot{\psi}}{3 \dot{\psi}^2} \, ,
\end{equation}
which simplifies to,
\begin{equation}
\dot{H}_{\mathrm{infl}} \, = \, - \dot{\psi}^2- \frac{2 \, U}{3}- U' -\frac{U''}{3} \,.
\label{hd}
\end{equation}

We now examine the above condition  for the  potential $U(\psi)$, for which several forms are existing \citep{Linde95,Linde05}. 
Let us consider a  typical well-known exponential form used in analytical treatment of the  inflation \citep{Weinberg08},
\begin{equation}
U(\psi) \, = \, g \, e^{ \mu \, \psi}\, ,      
\label{V}
\end{equation}
where $g$ and $\mu$ are constant, $g$ being positive  and $\mu$  generally negative.
The product $\mid \mu \,\psi \mid $ needs to be much smaller than 1, as required by   the condition
that the potential $U(\psi)$  does not vary too much over the inflation period, 
during which the time  $t $ varies by less than   $10^{-32}$ s. Moreover, we recall   that $\psi = \ln (t)$,
so that even if $t$ changes by a few powers of 10, $\psi$ changes only by a few unities.
 For now, we have: 
\begin{equation} 
U'(\psi) \, = \,g \, \mu \,  e^{\mu \, \psi}   \, , \quad \mathrm{and} \; \; U'' \, = \, g \, \mu^2 \,  e^{\mu \, \psi}\,.
\label{dv1}
\end{equation}
Then  relation (\ref{psit}) between and $\psi$ and the time $t$ gives
$ e^{\mu \, \psi} \, = \, e^{\mu \, \ln t} \, = \, t^{\mu} $, and therefore: 
\begin{equation}
U  = g \, t^{\mu}\,,  \; \;  U'  =  g \, \mu \, t^{\mu} \quad \mathrm{and} \; \; U''= g \, \mu^2 \, t^{\mu}
\end{equation}
With the above expressions for the potential and its derivatives, the expansion rate (\ref{hh}) becomes,
\begin{equation}
H_{\mathrm{infl}} \ = \, \frac{1}{t } -\frac{(2+ \mu) \, g}{3}\, t^{\mu+1}\,.
\label{hg}
\end{equation}
At inflation, the time  $t$,  expressed in the scale where  the present time  $t_0=1$,  
is a very small value of the order of  $4.4 \cdot 10^{-49}$. For values of $\mu > -1$, 
the second term on the right in the expression of $H_{\mathrm{infl}}$ would vanish near the origin. 
Thus, only the first term in Eqs. (\ref{hg}) would be significant. 
For $\mu=-1$, this second term would be a negative constant of order $-g/3$, it would contribute negatively, 
but will be dominated by the first term $1/t$ for sufficiently small $t$ near zero. 
For $\mu=-2$, the second term  would vanish, while  for $\mu$ more negative than -2, it contributes positively  and dominates. 
For the time derivative, we have from Eq. (\ref{hd}),
\begin{equation}
\dot{H}_{\mathrm{infl}} \, = \, -\frac{1}{t^2} -\frac{(\mu+2)(\mu+1) \, g}{3} \,t^{\mu}\,.
\label{dhg}
\end{equation}
There, the same kind of  remarks may be made. For $\mu$ more negative than -2, the second term on the right dominates. 
It is also negative and therefore provides graceful exit from inflation
due to its slowing down effect.

Let us consider negative values of $\mu < -2 $, which  correspond to  a typical exponentially decreasing potential   
in  inflation \citep{Weinberg08}. As seen above, the second terms in both $H_{\mathrm{infl}}$ and its derivative
dominate.
The critical ratio  (\ref{cond}) for the occurence of inflation becomes with
(\ref{hg}) and (\ref{dhg}),
\begin{equation}
\frac{\mid \dot{H}_{\mathrm{infl}} \mid }{H_{\mathrm{infl}}^2}\, = \, \frac{3\,(\mu+1)}{g \, (\mu+2)} \, t^{-\mu-2}\,.
\label{crit}
\end{equation}
For example, for  the case $\mu=-4$, we would get 
\begin{equation} 
H_{\mathrm{infl}}\approx  \frac{2 \,g}{3\,t^3}\,, \; \;   \mid \dot{H}_{\mathrm{infl}} \mid  \approx \frac{2\,g}{ t^4}\,, 
\; \; \;\; \mathrm{thus}, \; \; \frac{\mid \dot{H}_{\mathrm{infl}} \mid }{H_{\mathrm{infl}}^2}\, \approx \frac{9 t^2}{2 g} \, \ll \, 1\,.
\end{equation} 
The condition is satisfied since  the potential  $U(0)=g$ 
near the origin is  a finite quantity while the time becomes vanishingly small.  
Thus, the
{\emph{ relative}} variation of $H_{\mathrm{infl}}$ over a time $1/H_{\mathrm{infl}}$ is   very small. 
This  ensures a de Sitter-like exponential growth of the ``radius''  $a(t)$  of the Universe. 
Notice that after a time $t\approx \sqrt{2g/9}$ the inflation is smoothly over.

Let us consider that inflation starts at an  initial time $t_1$, larger than the Planck time, and is ending at time $t_2$,
which marks the beginning of the so-called phase of reheating, 
where the energy-density of the potential $U(\psi)$ starts being turned to baryogenesis. We have,
\begin{eqnarray}
\frac{a(t_2)}{a(t_1)}\, = \, \exp\left[\int^{t_2}_{t_1} H_{\mathrm{infl}} \, dt\right] \, = \, 
\exp\left[- \int^{t_2}_{t_1}  g \, \frac{(\mu+2)}{3} \, t^{\mu+1} dt\right] \, =  \nonumber \\
\, \exp \left[\,  \frac{g}{3}\, \,\left( t^{\mu+2}_1- t^{\mu+2}_2\right) \,\right] \,.  \quad
\label{integ}
\end{eqnarray}
For example, with $\mu = -4$, we would have
\begin{equation}
\frac{a(t_2)}{a(t_1)}\, = \exp \left[\,  \frac{g}{3}\, \left(\frac{1}{t^2_1}  -  \frac{1}{t^2_2} \right)\right] \,.
\end{equation}
The constant $g$, which represents the maximum of the potential $U$,  has  a large value  in current inflationary models. 
Interestingly enough, the way the time intervenes in Eq. (\ref{integ}), in $t^{\mu+2}$, 
produces a strong multiplication factor of the effect of the time interval for $\mu$-values  $\mu \leq -3$.  
Thus, depending on the $\mu$ value, it would even not be necessary 
that the potential $U$  is large to ensure a large number of e-foldings during inflation. 
This is an interesting  feature of inflation in the context of scale invariant models with a potential of the form 
$ U(\psi) = e^{\mu \psi}$. Another specific feature is that the field $\psi$, the inflaton, is related to time by $\psi = \ln t$, 
which ensures that the model Universe is slowly  ``rolling down'' the potential.
In phases following the inflation,  when $V(\varphi)$ and $U(\psi)$ starts changing significantly, the inflation comes to an end by several 
possibilities \citep{Linde95,Linde05}. At some stage, the energy-density of the potential $C \,U(\psi)$  is  turned into the 
baryogenesis by  the decay of supermassive Grand Unified Theory (GUT) 
bosons  producing the baryon asymmetry \citep{Kolb96,Linde96}.

The above developments show that the scale invariant equations do not substantially modify the
occurrence of the inflation, even if the usual scalar field $\varphi$ has been replaced  by 
the scalar field $\psi$, with which the energy density of the ordinary scale invariant vacuum is expressed. 
Although the scale invariant cosmological equations and the energy conservation
contain some additional terms, an exponential growth may be ensured with a high number of e-foldings
followed by a graceful exit from inflation.

Coming back to our main concern about the validity domain of scale invariance, we can say that  
scale invariance does not  prevent the existence of inflation.
Whether an inflation compatible with the equations of scale invariance permits scale invariance in the resulting
Universe remains an open question.
There is certainly room for philosophical discussions
as to whether scale invariance was present or not in the initial vacuum state before inflation. The only thing we 
may confirm here is that the mechanism of inflation could also be working in the scale invariant context.


Thus, the above study  does not  lead to a change of the previous conclusions we got from the discussion of the physical horizon. Scale invariance is forbidden, or at list has unclear yet interpretation, 
in cosmological models with $\Omega_{\mathrm{m}} \geq 1$. 
However, this may be also the key to reconcile Conformal Cyclic Cosmology 
\citep{Penrose:2006zz,Penrose'12} and Black-Hole Cosmology 
\citep{Pathria72, Dymnikova et al. 2001, Poplawski16, OshitaYokoyama18}
by providing the needed high potential energy-density $V(\varphi)$ via the confined central singularity of a black hole.
The question of scale invariance at low cosmological densities ($\Omega_{\mathrm{m}} < 1$) 
is an open possibility.

The answer may come from observations, in particular at large redshifts where 
significant differences with the $\Lambda$CDM models are predicted.
Comparisons  have been performed in the $m-z$ diagram for  the supernova data by \citet{Betoule14} up to redshift $z=1$,
there the differences  between scale invariant and $\Lambda$CDM  models are too small
\citep{Maeder17a,MaedGueor20a}. Another analysis has been made with the data by \citet{Lusso19}, 
where SN observations have been extended by data from quasars and gamma-ray bursts up to redshift $z=7$.
There some differences appear between the $\Lambda$CDM and scale invariant models. The scatter of the data prevents
a conclusion, but the situation is not far from allowing a decision in near future \citep{MaedGueor20a}.

\section{Conclusions} \label{concl}

We have shown on the basis of the scale invariant 
equations that scale invariance is  
forbidden, or at list has unclear yet interpretation,
for cosmological models with a mean density 
equal or above the critical density $\varrho_{\mathrm{c}}$.
The absence of scale invariance above $\varrho_{\mathrm{c}}$ is confirmed  by the fact that in this case
the  Universe models  have their mass-energy  distribution entirely contained within the limiting radius 
$R_{\mathrm{lim}}$, which may coincide with the Schwarzschild radius,
for causal connection. 

On the contrary, models with densities lower than $\varrho_{\mathrm{c}}$ have cosmological solutions according to the general 
scale invariant field equations by \citet{Dirac73} and \citet{Canuto77}.    The resulting models suggest  that 
the effects of scale invariance undergo a fast   decline from density $\varrho=0$ to $\varrho_{\mathrm{c}}$.
Such low density models have
some of their parts not causally connected. This lets open the possibility  that scale invariance is present. However, 
the answer is also related
to the cosmological history of the Universe, and in particular to the inflation. 

We have examined the occurrence of the inflation in the scale invariant context 
by identifying the scalar field $\varphi$ of the inflation with the scalar field $\psi$ associated to the energy-density of
the scale invariant space.
This identification also solves the so-called ``cosmological constant problem''.
Due to the properties of $\psi$, this leads to some change in the equations. 
For some forms of the potential $U(\psi)$, the scale invariant inflation is even favored with a high number of e-foldings. 
We conclude that the occurrence of scale invariance for models with 
$\Omega_{\mathrm{m}} \leq 1$  remains an open possibility, the answer of
which may come from high redshift observations.\\

\section*{Acknowledgments}
A.M. expresses his gratitude to his wife for her patience and support.
V.G. is extremely grateful to his wife and daughters for their understanding and family support  
during the various stages of the research presented.
Both authors are grateful to unknown referee for her/his 
useful remarks, among other ones for the most general solution for the scale factor given by Eq. (\ref{app}).
This research did not receive any specific grant from funding agencies in the public, commercial, or not-for-profit sectors.

\section*{Data availability}
No new data were generated or analyzed in support of this research.

\section{APPENDIX A:  The gauge factor $\lambda$.}

The most general solution for  $\lambda$ with $\lambda(t_0)=1$ has the form,
\begin{equation}
\lambda = \frac{1}{1+\kappa_0 \,(t-t_0)} \,,
\label{app}
\end{equation}
with $\kappa_0$ given by the expression:
\begin{equation}
\kappa_0 = -\left(\frac{1}{\lambda} \frac{d \lambda}{dt}\right)_0 \,.
\label{kappa0}
\end{equation}
This solution and the one given by Eq. (\ref{lt}) are equivalent.
We had searched for a solution of Eq. (\ref{diff1}) of the the general  form \citep{Maeder17a},
\begin{equation}
\lambda=a(t-b)^n + d \,.
\end{equation}
The equation  (\ref{diff1}) imposes $d=0$ and $n=-1$. Thus, we have
\begin{equation}
\lambda = \frac{a}{t-b}\,.
\end{equation}
The above  solution (\ref{app}) can also be written as
\begin{equation}
\lambda = \frac{1}{ \kappa_0\left[t -(t_0 -\frac{1}{\kappa_0})\right] }\,.
\label{app2}
\end{equation}
Thus, we have the correspondences $a = \frac{1}{ \kappa_0}$  and   $b  =  (t_0-\frac{1}{\kappa_0})$.
As mentioned in \citet{Maeder17a}, we may choose $a$, while any value of $b$ will satisfy the equations.
We note that if  $\lambda=t_0/t$  and $t_0=1$ we get with the expressions (\ref{app}) and (\ref{kappa0})
\begin{equation}
\kappa_0 = -\left(\frac{1}{\lambda} \frac{d \lambda}{dt}\right)_0 = t_0 (1/t_0^2)=1/t_0 \equiv  1 \,.
\end{equation}
Thus, we verify that  this is consistent with $a=1$  and $b=0$.  This means that the solution (\ref{lt})
and the one given by Eqs. (\ref{app}) and (\ref{app2}) are identical.  \\

\section{APPENDIX B:  The definition of the vacuum density}

In the usual  cosmological equations from General Relativity, 
\begin{eqnarray}
\frac{8 \, \pi G \varrho' }{3} = \frac{k}{a^2}+\frac{\dot{a}^2}{a^2}-\frac{\Lambda_E}{3} \, ,
\label{E1c} \\
-8 \, \pi G p'  = \frac{k}{a^2}+ 2 \frac{\ddot{a}}{a}+\frac{\dot{a}^2}{a^2}
-\Lambda_E  \, ,
\label{E2c}
\end{eqnarray}
 the cosmological constant $\Lambda_{\mathrm{E}}$ expressing 
the  energy density and pressure of the vacuum,   explicitly appear
 in the equations as  an additive term with the appropriate sign. It is important  to emphasize that $\varrho'$ and $p'$ are the density and pressure without the vacuum contribution. This leads to the identification,
\begin{equation}
\Lambda_{\mathrm{E}} \,= \,  8\,  \pi \,G \,\,\varrho'_{\mathrm{vac}} \,, \quad \mathrm{and} \; \;
 p'_{\mathrm{vac}}=-\varrho'_{\mathrm{vac}} \, ,
\label{defrp}
\end{equation}
where $p_{\mathrm{vac}}'$ and $\rho_{\mathrm{vac}}'$ are the pressure and density of the vacuum, while $p'$
and $\varrho'$ refer here to the matter and relativistic contributions.
Thus, the equation of state is verified.
In the scale invariant forms, the density $\varrho_{\mathrm{vac}}$,  pressure $p_{\mathrm{vac}}$  and $\Lambda_{\mathrm{E}}$
are the same as in GR, but multiplied by $\lambda^2$ \citep{Canuto77}.

Now, if we consider the scale invariant equations  (\ref{E1}) and (\ref{E2}) 
(obtained with the assumption that the vacuum space is scale invariant),
the vacuum effects expressed by $\lambda$ and its derivative
now only appear as  multiplicative factor of the expansion rate $H=\dot{a}/a$. 
Thus, the last terms of these equations cannot be identified with the vacuum density or pressure, 
since they are containing effects from both
the vacuum and the dynamical expansion.  

As an illustration, let us take $k=0 $ and write Eq.(\ref{E1}) as,
 \begin{equation}
H^2= \frac{8 \pi G}{3} \varrho - 2 \, H\frac{ \dot{\lambda}}{ \lambda} \, . 
\label{E1h}
\end{equation} 
If we would  interpret the last term in this equation as
 the vacuum density, we would get, 
\begin{equation}
\varrho_{\mathrm{vac}}= -\frac{3}{4 \pi G}  H \frac{\dot{\lambda}}{\lambda} \, .
\label{fx}
\end{equation}
which, with $H=-2 \dot{\lambda}/{\lambda}=2/t$,  is four times the (true) one given by (\ref{rhsiv}).
Proceeding similarly, we get $p_{\mathrm{vac}}$ from Eq. (\ref{E2}),
\begin{equation}
p_{\mathrm{vac}}=  \frac{4 \, H}{8 \pi G} \frac{\dot{\lambda}}{\lambda}=
\frac{ H}{2 \pi G} \frac{\dot{\lambda}}{\lambda} =  - \frac{2}{3} \varrho_{\mathrm{vac}}.
\label{fy}
\end{equation}
Thus, this process is not consistent with the equation of state of the vacuum, a situation which results
from the fact that the last terms in  Eqs.(\ref{E1}) and  (\ref{E2}) are not just the density and pressure of the vacuum.
Thus, we  should keep the direct estimate given by Eq. (\ref{rhsiv}),
\begin{equation} 
\varrho_{\mathrm{vac}}= \frac{3}{8 \pi G}  \frac{\dot{\lambda}^2}{\lambda^2} \, ,
\end{equation}
which verifies the equation of the vacuum state,
and does not combine  it with the energy of dynamical effects.

\bsp	
\label{lastpage}
\end{document}